\documentclass[fleqn,usenatbib]{mnras}

\usepackage{newtxtext,newtxmath}

\usepackage[T1]{fontenc}

\usepackage{graphicx}	
\usepackage{amsmath}	
\usepackage{multicol}        
\usepackage{bm}

\title[21-cm skew spectrum and smoothed skewness]{Skew spectrum and smoothed skewness of 21-cm signals from epoch of reionization}

\author[Ma et al.]{Qing-Bo Ma$^{1,2}$,\thanks{E-mail: \href{mailto:maqb@gznu.edu.cn}{maqb@gznu.edu.cn}}
Ling Peng $^{1}$\\
$^{1}$School of Physics and Electronic Science, Guizhou Normal University, Guiyang 550001, PR China \\
$^{2}$Guizhou Provincial Key Laboratory of Radio Astronomy and Data Processing, Guizhou Normal University, Guiyang 550001, PR China
}

\pubyear{2022}

\begin{document}
\label{firstpage}
\pagerange{\pageref{firstpage}--\pageref{lastpage}}
\maketitle

\begin{abstract}
Due to the non-linear ionizing and heating processes, the 21-cm signals from epoch of reionization (EoR) are expected to have strong non-Gaussian fluctuations. 
In this paper, we use the semi-numerical simulations to study the non-Gaussian statistics i.e. skew spectrum and smoothed skewness of the 21-cm signals from EoR. 
We find the 21-cm skew spectrum and smoothed skewness have similar evolution features with the 21-cm bispectrum. 
All of them are sensitive to the EoR models, while not too much to the cosmic volume applied. 
With the SKA1-low telescope as reference, we find both the skew spectrum and smoothed skewness have much higher S/N ratios than the 21-cm bispectrum.
\end{abstract}




\section{Introduction}
Following the first galaxies and first stars formation, the Universe undergoes the phase transition from fully neutral to highly ionized \citep{Furlanetto2006, Dayal2018}, which is named as the Epoch of Reionization (EoR).
Although many facilities have measured the properties about EoR, e.g. the optical depth measured by the Cosmic Microwave Background (CMB) projects \cite[$\tau= 0.0544\pm0.007$ by ][]{Planck2020} denotes that the mid-point redshift of the EoR is at $z=7.68\pm 0.79$, the Gunn-Peterson trough measured by the spectra of high-$z$ quasars \citep{Fan2006} confirms the end of EoR at $z \sim 6$, and many high-$z$ galaxies during EoR have been observed by the Hubble Space Telescope (HST) \citep{Bouwens2015} and James Webb Space Telescope (JWST) \citep{Donnan2022}, the most promising method is to measure the 21-cm signals of the hyper-fine transition line of neutral hydrogen \citep{Furlanetto2006, Koopmans2015}.
Indeed, the measurements of 21-cm signals, including the global and the interference ones, have been one of the main goals of many radio telescopes, e.g. the Experiment to Detect the Global EoR Signature \cite[EDGES, ][]{Bowman2018}, the Shaped Antenna measurement of the background Radio Spectrum 3 telescope \cite[SARAS-3, ][]{Bevins2022}, the Low-Frequency Array \cite[LOFAR, ][]{Mertens2020}, the Murchison Widefield Array \cite[MWA, ][]{Trott2020}, the Hydrogen Epoch of Reionization Array \cite[HERA, ][]{Abdurashidova2022}, and the Square Kilometre Array \cite[SKA, ][]{Koopmans2015}. 

Although without clearly definite results reported by the 21-cm facilities, some telescopes have released the early results of 21-cm signal, e.g. the EDGES project has reported an absorption profile on the global 21-cm signal at 78~MHz (i.e. $z\sim 17$) \citep{Bowman2018}, while this is still debated \cite[e.g. ][]{Hills2018, Singh2022}, and not confirmed by the recent observations of SARAS-3 telescope \citep{Bevins2022}.
The interference telescopes have given some upper limits on the 21-cm power spectra $\Delta_{\rm 21cm}^{2}$, e.g. a 2-$\sigma$ upper limit of $\Delta_{\rm 21cm}^{2} < (73)^{2}\,\rm mK^{2}$ at $z \approx 9.1$ and $k=0.075\,h\,\rm cMpc^{-1}$ by the LOFAR telescope \citep{Mertens2020}, $\Delta_{\rm 21cm}^{2} \le (43)^{2}\,\rm mK^{2}$ at $z=6.5$ and $k=0.14\,h\,\rm cMpc^{-1}$ by the MWA telescope \citep{Trott2020}, and $\Delta_{\rm 21cm}^{2} \le (30.76)^{2}\,\rm mK^{2}$ at $z=7.9$ and $k=0.192\,h\,\rm cMpc^{-1}$ by the HERA telescope \citep{Abdurashidova2022}.
With these results, people already rule out some extreme EoR models \cite[e.g. ][]{Ghara2020, Ghara2021, Greig2021MWA, Greig2021LOFAR, Abdurashidova2022b}.

Since the ionizing and heating processes during EoR are highly non-linear, the fluctuations on the 21-cm images are very non-Gaussian \citep{Majumdar2018}, which can be measured by e.g. the skewness and kurtosis \citep{Kittiwisit2022}, the three point correlation function \citep{Hoffmann2019}, the bispectrum \citep{Watkinson2019, Hutter2020, Ma2021} and bispectrum phase \citep{Thyagarajan2020}, and the position-dependent power spectra \citep{Giri2019}. 
Although these high-order statistics may be hard to measure \citep{Watkinson2021}, they are very sensitive to the physical processes of EoR e.g. the reionization history \citep{Majumdar2018}, the X-ray heating \citep{Watkinson2019, Ma2021}, the ionization topologies \citep{Hutter2020}, and the effect of redshift space distortion \citep{Majumdar2020}.
Meanwhile, their combination with 21-cm power spectra observations will improve the constraints on the parameters of EoR model \citep{Shimabukuro2017}.

The calculations of non-Gaussian features (e.g. the three point correlation function and the bispectrum) are usually very expensive both for simulations and observational datas, even with the FFT-based technique developed by \cite{Watkinson2017} to compute the 21-cm bispectrum, while some quantities are easier to compute and can denote the similar features, e.g. the skew spectrum that used to describe the non-Gaussian features of CMB \citep{Cooray2001} and galaxy distribution \citep{Moradinezhad2020, Dai2020b}.
The computing of skewness is also very convenient, while there is only one value for each 21-cm image.
However, after smoothing the smaller scale fluctuations with different $k$s, the 21-cm skewness (called as smoothed skewness) shows similar behaviour with the 21-cm bispectrum \citep{Ma2021}.
In this paper, we will study how the skew spectrum and smoothed skewness of 21-cm signals from EoR evolve with redshift, their relation with the reionization history, and their detectability by the 21-cm telescope SKA1-low. 
We will also compare the results with those of 21-cm bispectrum.

The following paper is organized as: the simulations and the methods to computer bispectrum, skew spectrum and smoothed skewness are described in Section~\ref{sec:method}, the results are in Section~\ref{sec:res},  the conclusions are summarized in Section~\ref{sec:concl}.
The adopted cosmological parameters are $\Omega_{\Lambda} = 0.685$, $\Omega_{m} = 0.315$, $\Omega_{b} = 0.0493$, $\sigma_{8} = 0.811$, $n_{s} = 0.965$ and $h = 0.674$ \citep{Planck2020}.

\section{Method}
\label{sec:method}
We describe the simulations adopted in Sec.~\ref{sec:method:simul}, and then the methods to compute the bispectrum in Sec.~\ref{sec:method:bis}, the skew spectrum in Sec.~\ref{sec:method:skewspe}, and the smoothed skewness in Sec.~\ref{sec:method:skew}.

\subsection{Simulations}
\label{sec:method:simul}
We use the semi-numerical simulation 21CMFAST \citep{Mesinger2011} to mimic the evolution of matter density, ionization and temperature status of gas medium, and the 21-cm differential brightness temperature (DBT, $\delta T_{\rm 21cm}$).
The simulations start at $z=35$ and end at $z=6$, with 57 snapshot outputs from $z=20$ to $6$.
The fiducial simulation (named as L600) has box length $600\,\rm cMpc$ and grid number $800^3$. 
The number of ionizing photons per stellar baryon adopted is $N_{\rm UV} = 5000$.
The fraction of collapsed gas that form stars is $f_{\ast} = f_{\ast,\,0} \times (M_{h}/10^{10} {\rm M}_{\odot})^{\alpha_{\ast}}$, where $M_{h}$ is the halo mass, $f_{\ast,\,0} = 0.05$ and $\alpha_{\ast} = 0.5$.
The escape fraction is $ f_{\rm esc} = f_{\rm esc,\,0} \times (M_{h}/10^{10} {\rm M}_{\odot})^{\alpha_{\rm esc}}$ \citep{Park2019}, where $f_{\rm esc,\,0} = 0.1$ and $\alpha_{\rm esc} = -0.5$.
The number of halos hosting active star-forming galaxies (i.e. the duty cycle) is $f_{\rm duty} \propto {\rm exp}({-M_{\rm turnover}/M_{h}})$ \citep{Greig2022}, where the halo mass threshold for efficient star formation $M_{\rm turnover} = 5\times 10^{8}\,\rm M_{\odot}$.
The SED of X-ray binary is from \citet{Fragos2013b}, with the luminosity $L_{<2 {\rm keV}}/{\rm SFR} = 10^{40.5} \rm erg\, s^{-1}\, M_{\odot}^{-1}\, yr$.

\begin{figure}
\centering
\includegraphics[width=0.95\linewidth]{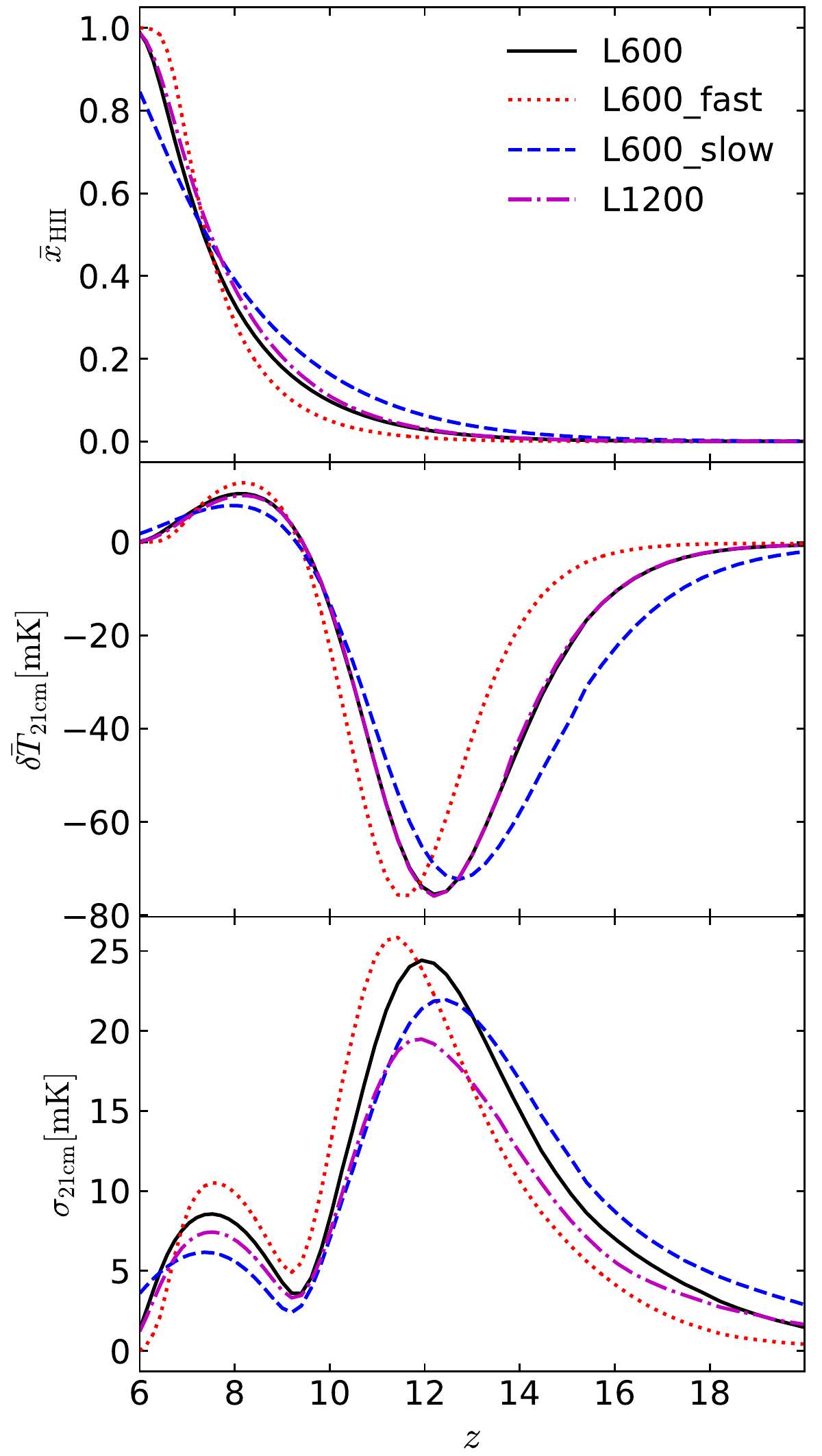}
  \caption{History of volume averaged ionization fraction $\bar{x}_{\rm HII}$ (top), global 21-cm DBT $\bar{\delta T}_{\rm 21cm}$ (central) and standard deviation $\sigma_{\rm 21cm}$ of $\delta T_{\rm 21cm}$ (bottom) from simulations L600 (solid black), L600\_fast (dotted red), L600\_slow (dashed blue) and L1200 (dash-dotted magenta).
  }
\label{fig:0d_statis}
\end{figure}

We also run three more simulations as comparisons. 
The first one is with faster ionization process (named as L600\_fast), which increases the star forming fraction $f_{\ast,\,0}$ and the halo mass threshold $M_{\rm turnover}$ for efficient star formation, i.e. with $f_{\ast,\,0} = 0.15$ and $M_{\rm turnover} = 5\times 10^{9}\,\rm M_{\odot}$.
The second one is with slower ionization process (named as L600\_slow), i.e. with $f_{\ast,\,0} = 0.024$ and $M_{\rm turnover} = 5\times 10^{7}\,\rm M_{\odot}$. 
Note that in these two simulations, we increase/decrease the halo mass threshold $M_{\rm turnover}$ with one magnitude, while fine-tune the parameter $f_{\ast,\,0}$ to make sure the simulations have the consistent half-ionization redshifts with L600 (see Fig.~\ref{fig:0d_statis}).
The third one is with box length $1200\,\rm cMpc$ and grid number $800^3$ (named as L1200), and the same parameter values of EoR model with L600. 
Compared to L600, L1200 covers larger cosmic volume but with lower spatial resolution.

As a reference, Fig.~\ref{fig:0d_statis} shows the evolution history of volume averaged ionization fraction $\bar{x}_{\rm HII}$, global mean $\bar{\delta T}_{\rm 21cm}$ and standard deviation $\sigma_{\rm 21cm}$ of $\delta T_{\rm 21cm}$, from four simulations.
These simulations have $\bar{x}_{\rm HII} = 0.5$ at $z \approx 7.4$, with the redshift gaps $\Delta z$ from $\bar{x}_{\rm HII} = 0.1$ to 0.9 being $\Delta z \approx 3.8$ for L600 and L1200, $\Delta z \approx 2.5$ for L600\_fast and $\Delta z >5$ for L600\_slow. 
The simulation L1200 has similar histories of $\bar{x}_{\rm HII}$ and $\bar{\delta T}_{\rm 21cm}$ with L600, while has smaller $\sigma_{\rm 21cm}$ than the latter, due to its lower spatial resolution.
Compared to L600, the faster ionization of L600\_fast results in higher peak of $\bar{\delta T}_{\rm 21cm}$ at $z \approx 8.2$, and higher peak of  $\sigma_{\rm 21cm}$ at $z \approx 7.5$. 
The higher halo mass threshold $M_{\rm turnover}$ delays the X-ray heating in L600\_fast, thus the absorption trough on $\bar{\delta T}_{\rm 21cm}$ is at $z \approx 11.5$, lower than that of L600 ($z \approx 12.2$), while it increases the peak of $\sigma_{\rm 21cm}$ in the early time of EoR ($z \approx 11.4$). 
Instead, the slower ionization of L600\_slow leads to lower peak on $\bar{\delta T}_{\rm 21cm}$ at $z \approx 8.2$ and $\sigma_{\rm 21cm}$ at $z \approx 7.5$.
The lower value of $M_{\rm turnover}$ in L600\_slow allows the star formation within halos with lower-limit mass lower than L600, and thus earlier X-ray heating, i.e. the earlier absorption trough on $\bar{\delta T}_{\rm 21cm}$ (at $z\approx 12.6$) than L600, while it reduces the peak of $\sigma_{\rm 21cm}$ at $z \approx 12.5$. 

\subsection{Bispectrum}
\label{sec:method:bis}
The bispectrum of 21-cm image ($b_{\rm 21cm}$) is the statistics of three point correlations of $\delta T_{\rm 21cm}$ in the Fourier space, which can be expressed as:
\begin{align}
b_{\rm 21cm}(\bm{k_1},\bm{k_2},\bm{k_3}) =  &  \delta_{D}(\bm{k_1}+\bm{k_2}+\bm{k_3} )  \notag\\
&\times \langle \widetilde{\delta T}_\mathrm{21cm}(\bm{k_1}) \widetilde{\delta T}_\mathrm{21cm}(\bm{k_2}) \widetilde{\delta T}_\mathrm{21cm}(\bm{k_3})\rangle
\end{align}
where $\widetilde{\delta T}_\mathrm{21cm}(\bm{k_{i}})$ ($i=1,2,3$) is the Fourier transform of $\delta T_\mathrm{21cm}$. 
With different lengths of $\bm{k_1}$, $\bm{k_2}$ and $\bm{k_3}$, i.e. $k_{1}$, $k_{2}$ and $k_{3}$, $b_{\rm 21cm}$ has many modes \cite[see e.g. ][]{Majumdar2020} that denote different non-Gaussian features. 
In this paper, we will focus only on the one of equilateral triangles i.e. $k_{1} = k_{2} = k_{3} = k$.

We use the fast Fourier transform (FFT) based technique \citep{Watkinson2017} to compute the $b_{\rm 21cm}$.
It is much faster than the traditional triangle counting technique, while gives the consistent $b_{\rm 21cm}$ results \citep{Watkinson2017}. 
We also normalize the 21-cm bispectrum as $B_{\rm 21cm} (k) =b_{\rm 21cm} (k) \times k^{6}/(2\pi^{2})^{2}$ \citep{Ma2021} in the following. 

\subsection{Skew spectrum}
\label{sec:method:skewspe}
Following the definition of skew spectrum for the CMB \citep{Cooray2001} and the galaxy distribution \citep{Dai2020b, Dai2020}, the skew spectrum of 21-cm images ($s_{\rm 21cm}$) is defined as the cross-spectrum of $\Phi_{T_{\rm 21cm}}= (\delta T_{\rm 21cm} - \bar{\delta T}_{\rm 21cm})^{2}$ with $\delta T_{\rm 21cm}$, i.e.
\begin{equation}
    s_{\rm 21cm} (\bm{k}) = \delta_{D}(\bm{k}+\bm{k'}) 
 \times \langle \widetilde{\Phi}_ {T_\mathrm{21cm}}(\bm{k}) \widetilde{\delta T}_\mathrm{21cm}(\bm{k'})\rangle
\end{equation}
where $\widetilde{\Phi}_{T_\mathrm{21cm}} (\bm{k})$ is the Fourier transform of $\Phi_{T_{\rm 21cm}}$. 

Since $s_{\rm 21cm}$ is the integration of the $b_{\rm 21cm}$ modes \cite[see the discussions in e.g.][]{Dai2020b}, it is expected to display the similar features to $b_{\rm 21cm}$.
As the computing of $s_{\rm 21cm}$ is actually the two points cross-correlation, it is much faster than $b_{\rm 21cm}$. 
To keep the same units with $B_{\rm 21cm}$, the 21-cm skew spectrum is normalized as $S_{\rm 21cm} (k) = s_{\rm 21cm} (k) \times k^{3}/2\pi^{2}$.

\subsection{Smoothed skewness}
\label{sec:method:skew}
The smoothed skewness ($\Gamma_{\rm 21cm}$) is defined as the skewness of the $\delta T_\mathrm{21cm}$ images after smoothing the smaller scale fluctuations with wave-number $>k$, i.e. only keep signals at larger scales than $k$s, which can be computed by:
\begin{equation}
    \Gamma_{\mathrm{21cm}} (k) = \frac{\sum (\hat{\delta T}_{\mathrm{21cm},k} - \bar{\delta T}_\mathrm{21cm})^{3}}{N}
\end{equation}
where $\hat{\delta T}_{\mathrm{21cm},k}$ denotes the $\delta T_\mathrm{21cm}$ images after smoothing the fluctuations with wave-number $>k$, $N$ is the cell number after smoothing.

Since the calculation of $\Gamma_{\rm 21cm}$ does not need Fourier transform, it is faster than $B_{\rm 21cm}$, while it loops the whole 21-cm images for each $k$, thus slower than $S_{\rm 21cm}$ to produce the full spectrum. 
As showed in \cite{Ma2021}, $\Gamma_{\rm 21cm}$ can presents similar non-Gaussian features with $B_{\rm 21cm}$.
Actually, $\Gamma_{\rm 21cm}$ is the integration of the $b_{\rm 21cm}$ modes \citep{Shimabukuro2016}:
\begin{equation}
    \Gamma_{\mathrm{21cm}} (k) = \int^{k} \frac{{\rm d}^{3} \bm{k}_{1}}{(2\pi)^{3}} \int^{k} \frac{{\rm d}^{3} \bm{k}_{2}}{(2\pi)^{3}} b_{\rm 21cm}(\bm{k_{1}}, \bm{k_{2}}, -\bm{k_{1}}-\bm{k_{2}}).
\end{equation}

\section{Results}
\label{sec:res}
We present the evolution features of $B_{\rm 21cm}$, $S_{\rm 21cm}$ and $\Gamma_{\rm 21cm}$ in Sec.~\ref{sec:res:evol}, their model dependence in Sec.~\ref{sec:res:model}, and their detectability in Sec.~\ref{sec:res:detec}.

\subsection{Evolution features}
\label{sec:res:evol}
\begin{figure}
\centering
\includegraphics[width=0.95\linewidth]{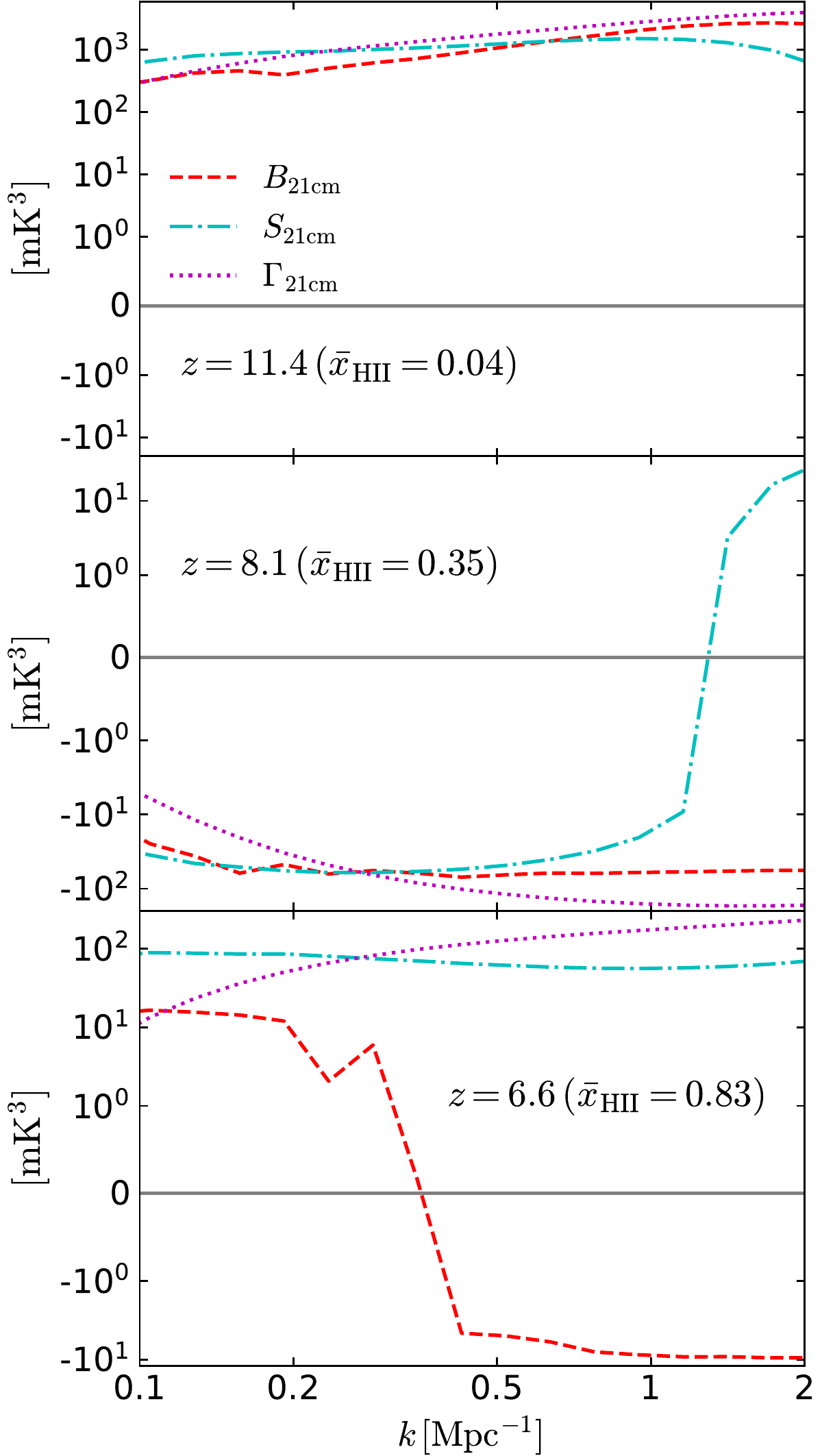}
  \caption{Bispectrum $B_{\rm 21cm}$ (dashed red), skew spectrum $S_{\rm 21cm}$ (dash-dotted cyan) and smoothed skewness $\Gamma_{\rm 21cm}$ (dotted magenta) of $\delta T_\mathrm{21cm}$ at $z= 11.4$ ($\bar{x}_{\rm HII} = 0.04$, top), 8.1 ($\bar{x}_{\rm HII} = 0.35$, central) and 6.6 ($\bar{x}_{\rm HII} = 0.83$, bottom), from the simulation L600.
  }
     \label{fig:bis_comp_3z}
\end{figure}
Fig.~\ref{fig:bis_comp_3z} shows the $B_{\rm 21cm}$, $S_{\rm 21cm}$ and $\Gamma_{\rm 21cm}$ from the simulation L600, at three $z$s. 
At the early stage of EoR, e.g. at $z=11.4$, that after X-ray heating (see Fig.~\ref{fig:0d_statis}), the non-Gaussianity of $\delta T_{\rm 21cm}$ is dominated by the matter density, thus $B_{\rm 21cm}$, $S_{\rm 21cm}$ and $\Gamma_{\rm 21cm}$ are all positive within $k=0.1-2\,\rm Mpc^{-1}$. 
With the proper normalization discussed in Sec.~\ref{sec:method}, they present similar amplitudes.
At $z=8.1$, the ionized bubbles starts to dominate the behaviour of non-Gaussian features, that leads to negative $B_{\rm 21cm}$ \citep{Majumdar2018}.
Consistently, the $S_{\rm 21cm}$ and $\Gamma_{\rm 21cm}$ are also negative, except that $S_{\rm 21cm}$ is positive at $k > 1.3\,\rm Mpc^{-1}$.
At the end stage of EoR, e.g. at $z=6.6$, the non-Gaussian features of $\delta T_{\rm 21cm}$ are dominated by the islands of neutral hydrogen, which results in some positive $B_{\rm 21cm}$ \citep{Majumdar2018}, e.g. at $k < 0.35 \,\rm Mpc^{-1}$.
Differently, the $S_{\rm 21cm}$ and $\Gamma_{\rm 21cm}$ are fully positive within $k=0.1-2\,\rm Mpc^{-1}$.
This is due to the fact that both the $S_{\rm 21cm}$ and $\Gamma_{\rm 21cm}$ are the integration of $B_{\rm 21cm}$ modes, i.e. the results of $S_{\rm 21cm}$ and $\Gamma_{\rm 21cm}$ at $k> 0.35 \,\rm Mpc^{-1}$ can include the contributions of $B_{\rm 21cm}$ modes from the scales with $k < 0.35 \,\rm Mpc^{-1}$. 

\begin{figure}
\centering
\includegraphics[width=0.95\linewidth]{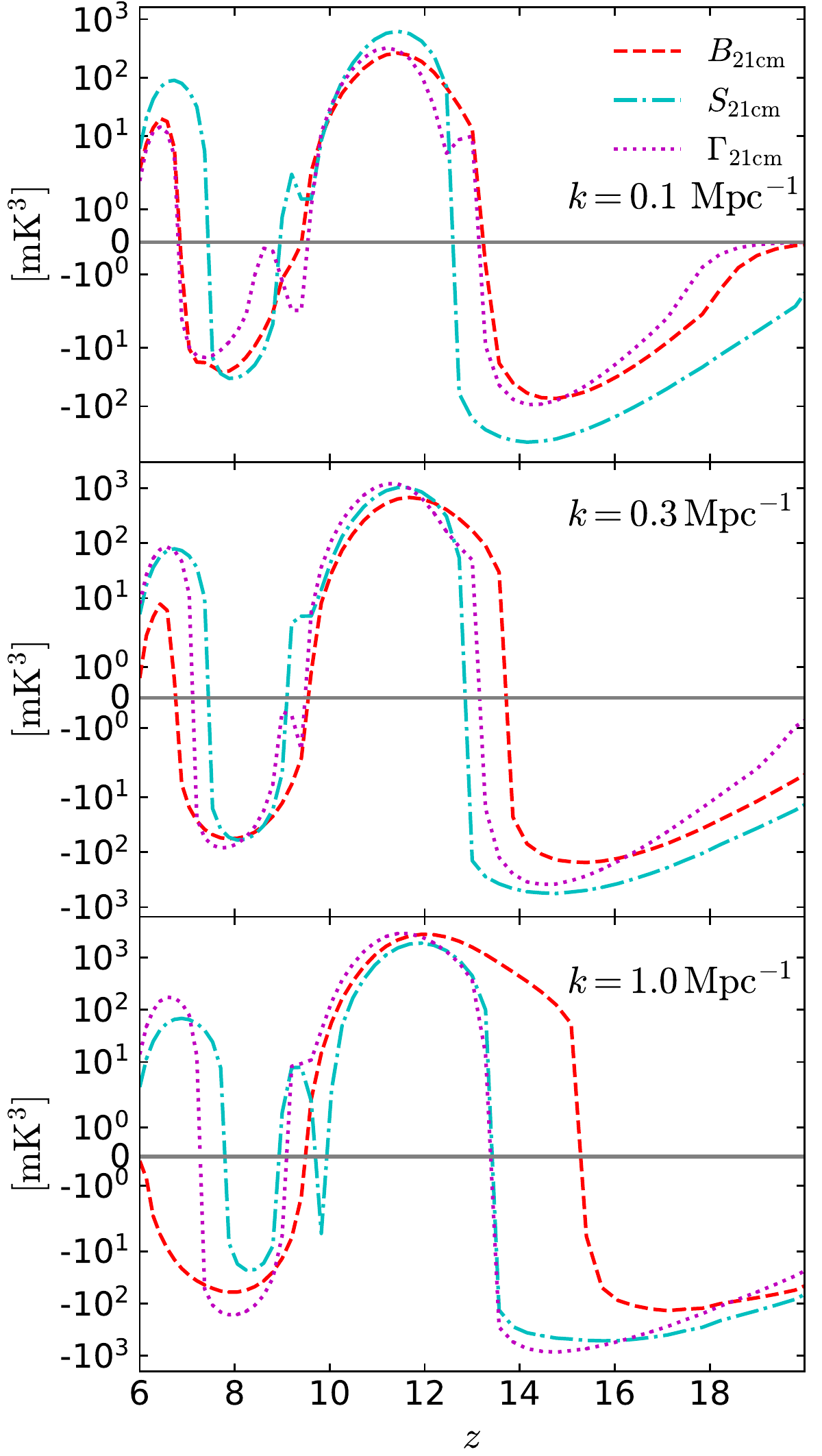}
  \caption{Evolution of $B_{\rm 21cm}$ (dashed red), $S_{\rm 21cm}$ (dash-dotted cyan) and $\Gamma_{\rm 21cm}$ (dotted magenta) with redshift, at $k=0.1\,\rm Mpc^{-1}$ (top), $k=0.3\,\rm Mpc^{-1}$ (central), and $k=1.0\,\rm Mpc^{-1}$ (bottom), from the simulation L600.
          }
     \label{fig:bis_comp_3k}
\end{figure}
Fig.~\ref{fig:bis_comp_3k} presents the redshift evolution of $B_{\rm 21cm}$, $S_{\rm 21cm}$ and $\Gamma_{\rm 21cm}$ from the simulation L600 at three $k$s.
They display similar evolution features at all three $k$s.
Specifically, all of them are negative at $z > 15$, i.e. in the phase of Ly$\alpha$ pumping that couples the spin temperature of neutral hydrogen to the kinetic temperature of gas medium.
When the X-ray radiation dominates the heating of IGM, i.e. within $z = [10, 12]$, they become positive, as the X-ray heating positively couples the spin temperature of neutral hydrogen to the matter density. 
With the ionization continuing, the ionized bubbles start to dominate the non-Gaussian features of $\delta T_{\rm 21cm}$ and lead to negative $B_{\rm 21cm}$, $S_{\rm 21cm}$ and $\Gamma_{\rm 21cm}$.
As mentioned earlier, at the end of EoR e.g. $z<7$, the non-Gaussian features of $\delta T_{\rm 21cm}$ are dominated by the neutral hydrogen islands, that results in positive $B_{\rm 21cm}$, $S_{\rm 21cm}$ and $\Gamma_{\rm 21cm}$.
However, some differences also appear at three $k$s.
Specifically, at $k=0.1\,\rm Mpc^{-1}$, the evolution of $\Gamma_{\rm 21cm}$ is similar to $B_{\rm 21cm}$, while $S_{\rm 21cm}$ shows higher absolute amplitude.
This is because the $\Gamma_{\rm 21cm}$ at $k=0.1\,\rm Mpc^{-1}$ is only from the $B_{\rm 21cm}$ models at $k \le 0.1\,\rm Mpc^{-1}$, thus show more similar features with $B_{\rm 21cm}$.
At $k=0.3\,\rm Mpc^{-1}$, the absolute amplitude of $\Gamma_{\rm 21cm}$ is higher than $B_{\rm 21cm}$, and it is closer to $S_{\rm 21cm}$, since both the $\Gamma_{\rm 21cm}$ and the $S_{\rm 21cm}$ at $k=0.3\,\rm Mpc^{-1}$ includes many $B_{\rm 21cm}$ modes, thus have higher absolute amplitude than $B_{\rm 21cm}$. 
At $k=1.0\,\rm Mpc^{-1}$, both $S_{\rm 21cm}$ and $\Gamma_{\rm 21cm}$ are positive at $z<8$, while $B_{\rm 21cm}$ is negative at the same $z$s. 
As mentioned before, this is because $S_{\rm 21cm}$ and $\Gamma_{\rm 21cm}$ are the integration of $B_{\rm 21cm}$ modes, i.e. they include the contributions from $B_{\rm 21cm}$ modes with smaller $k$s.

\subsection{Model dependence}
\label{sec:res:model}
\begin{figure*}
\centering
\includegraphics[width=0.95\linewidth]{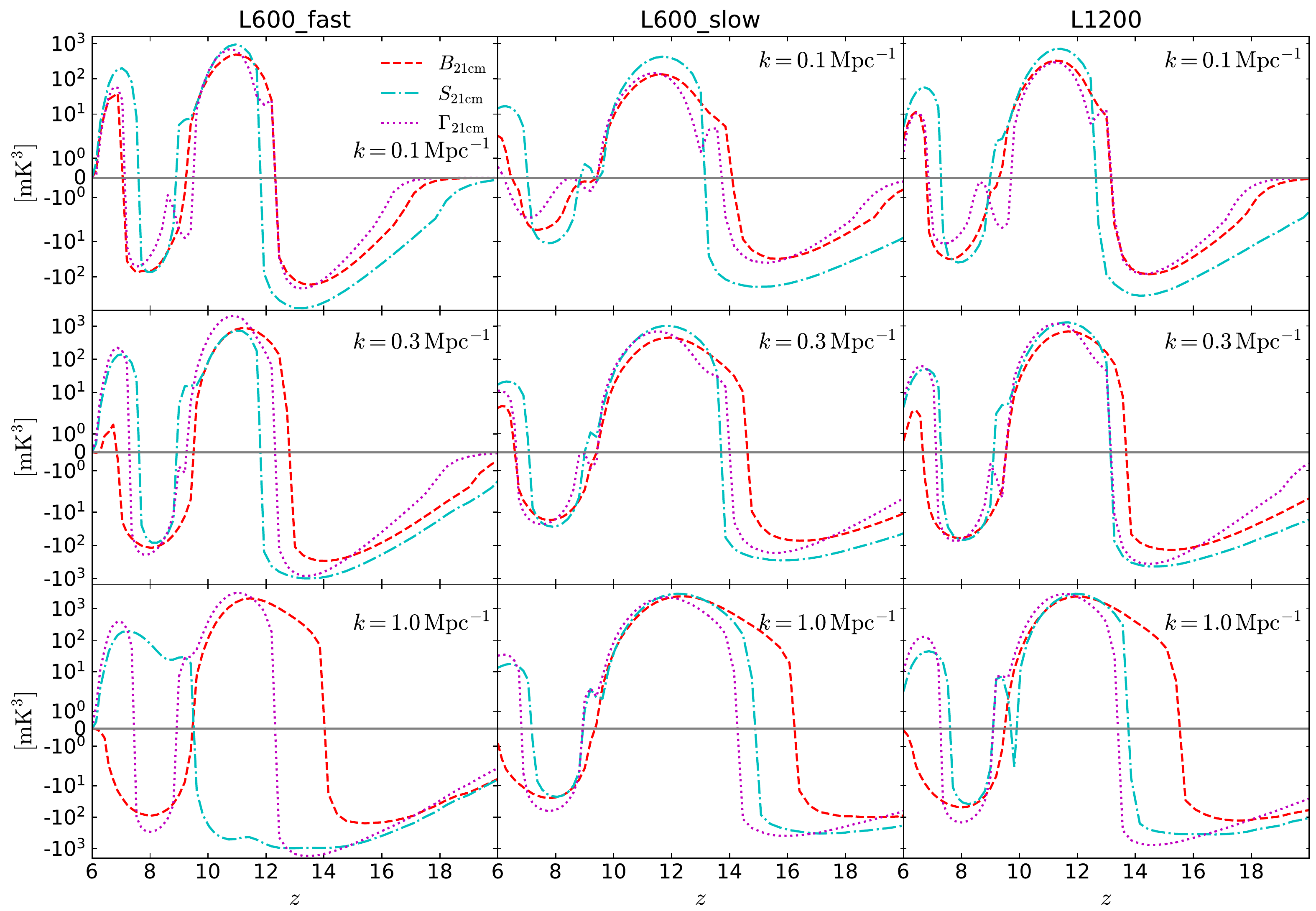}
  \caption{Evolution of $B_{\rm 21cm}$ (dashed red), $S_{\rm 21cm}$ (dash-dotted cyan) and $\Gamma_{\rm 21cm}$ (dotted magenta) at $k=0.1\,\rm Mpc^{-1}$ (top), $k=0.3\,\rm Mpc^{-1}$ (central), and $k=1.0\,\rm Mpc^{-1}$ (bottom) from simulation L600\_fast (left), L600\_slow (middle) and L1200 (right).}
     \label{fig:bis_comp_3k_diffM}
\end{figure*}
Fig.~\ref{fig:bis_comp_3k_diffM} shows the redshift evolution of $B_{\rm 21cm}$, $S_{\rm 21cm}$ and $\Gamma_{\rm 21cm}$ from simulations L600\_fast, L600\_slow and L1200.
As showed in Fig.~\ref{fig:0d_statis}, these three simulations have the same half-ionization $z$ with L600, while L600\_fast (L600\_slow) has faster (slower) ionization process, and L1200 has larger box length. 

With faster ionization, i.e. simulation L600\_fast, the 21-cm signals and fluctuations are more significant (see Fig.~\ref{fig:0d_statis}). 
Thus $B_{\rm 21cm}$, $S_{\rm 21cm}$ and $\Gamma_{\rm 21cm}$ present higher amplitudes than L600.
In L600\_fast, the evolution features of $B_{\rm 21cm}$, $S_{\rm 21cm}$ and $\Gamma_{\rm 21cm}$ are consistent at $k=0.1$ and 0.3 $\rm Mpc^{-1}$, while obviously different at $k=1.0\,\rm Mpc^{-1}$.
With slower ionization, i.e. simulation L600\_slow, the 21-cm signals and fluctuations are smaller (see Fig.~\ref{fig:0d_statis}). 
Thus the amplitudes of $B_{\rm 21cm}$, $S_{\rm 21cm}$ and $\Gamma_{\rm 21cm}$ are lower than L600.
At $k=0.1\,\rm Mpc^{-1}$, the common features of $B_{\rm 21cm}$, $S_{\rm 21cm}$ and $\Gamma_{\rm 21cm}$ are not significant like that in L600, e.g. the $\Gamma_{\rm 21cm}$ is obviously different to $B_{\rm 21cm}$ at $z<9$, 
while $B_{\rm 21cm}$, $S_{\rm 21cm}$ and $\Gamma_{\rm 21cm}$ have clearly consistent evolution at $k=0.3\rm Mpc^{-1}$ and $1.0 \, \rm Mpc^{-1}$ especially at $z>8$.
L1200 covers larger cosmic volume than L600, i.e. $S_{\rm 21cm}$ and $\Gamma_{\rm 21cm}$ include more large scale modes of $B_{\rm 21cm}$ than those in L600, while they display almost the same evolution with the latter.
This means that the simulations or 21-cm surveys in the near future with larger cosmic volume would not significantly affect the results of $B_{\rm 21cm}$,  $S_{\rm 21cm}$ and $\Gamma_{\rm 21cm}$. 

As a summary, with the different EoR models, $S_{\rm 21cm}$ and $\Gamma_{\rm 21cm}$ still present similar evolution features with $B_{\rm 21cm}$, and all of them are sensitive to the EoR models, while not too much to the cosmic volume applied. 

\subsection{Detectability}
\label{sec:res:detec}
We adopt SKA1-low telescope as the reference facility to predict the signal/noise ratios (S/N) of $B_{\rm 21cm}$, $S_{\rm 21cm}$ and $\Gamma_{\rm 21cm}$.
SKA1-low has $N_{\rm st} = 224$ stations in the compact core with diameter $D = 1\, \rm km$ to measure the 21-cm visibility from EoR. 
The root mean square (rms) of instrumental noise on the $\delta T_{\rm 21cm}$ images can be estimated by:
\begin{equation}
    \sigma_{N} = \frac{\lambda^{2}}{S \Omega_{\rm beam} \sqrt{N_{\rm st}(N_{\rm st} - 1) R_{\rm width} t_{\rm int}}}
\end{equation}
where $S$ is the telescope sensitivity from \cite{Dewdney2016}, the solid angle of beam $\Omega_{\rm beam} = 1.133\theta^{2}$, the angular resolution $\theta = \lambda/D$, and $\lambda = 21\,{\rm cm} \times (1+z)$.
We assume the spectral resolution $R_{\rm width} = 0.1\,\rm MHz$, and the integration time $t_{\rm int} = 1000\,\rm hr$.
We also assume that the total survey area of SKA1-low is $100\, {\rm deg}^{2}$, and the frequency band for each observation is $2\,\rm MHz$.

By assuming the instrumental noise is fully Gaussian, the bispectrum noise can be estimated by $b_{N} = \sigma_{N}^{3} x^{4} y^{2}$ \citep{Yoshiura2015}, where $x$ and $y$ are the comoving distances corresponding to the angular resolution $\theta$ and the frequency resolution $R_{\rm width}$. 
If including the sampling error, the expected S/N ratio of $B_{\rm 21cm}$ is
\begin{equation}
    {\rm \frac{S}{N}} = \frac{b_{\rm 21cm}(k)}{\sqrt{\left(b_{\rm 21cm}^{2}(k) +b_{N}^{2}\right)/N_{\rm tri}(k)}}
\end{equation}
where $N_{\rm tri}(k)$ is the triangle number of bispectrum mode with $k$ measured by SKA1-low \cite[see the formula in ][]{Yoshiura2015, Ma2021}.

The instrumental noise on skew spectrum is approximately computed by $s_{N} = \sigma_{N}^{3} x^{2} y$.
This is similar to the power spectrum \cite[see e.g.][]{Yoshiura2015}, while multiplying an extra $\sigma_{N}$.
The expected S/N of $S_{\rm 21cm}$ is then
\begin{equation}
    {\rm \frac{S}{N}} = \frac{s_{\rm 21cm}(k)}{\sqrt{\left(s_{\rm 21cm}^{2}(k) +s_{N}^{2}\right)/ N_{\rm pair}(k)}}
\end{equation}
where $N_{\rm pair}(k)$ is the number of power spectrum mode $k$ that will be measured by SKA1-low \cite[see e.g.][]{Yoshiura2015, Ma2021}. 

After removing the fluctuations with wave-number $>k$, the skewness noise of smoothed cells (with volume $V_{\rm cell} = (2\pi/k)^{3}$) is estimated as $\gamma_{N,\, \rm cell}(k) = (\sigma_{N}/\sqrt{N_{\rm pixel}(k)})^{3}$, where $N_{\rm pixel}(k)$ is the pixel number of measured 21-cm images within the cells with volume $V_{\rm cell}$.
The S/N ratio of $\Gamma_{\rm 21cm}$ is then 
\begin{equation}
    {\rm \frac{S}{N}} = \frac{\Gamma_{\rm 21cm}(k) }{\sqrt{\left(\Gamma_{\rm 21cm}^{2}(k) +\gamma_{N,\,\rm cell}^{2}(k)\right)/{N_{\rm cell}(k)}}}
\end{equation}
where $N_{\rm cell}(k)$ is the number of smoothed cells with volume $V_{\rm cell}$ within the cosmic volume measured by SKA1-low telescope. 

\begin{figure}
\centering
\includegraphics[width=0.95\linewidth]{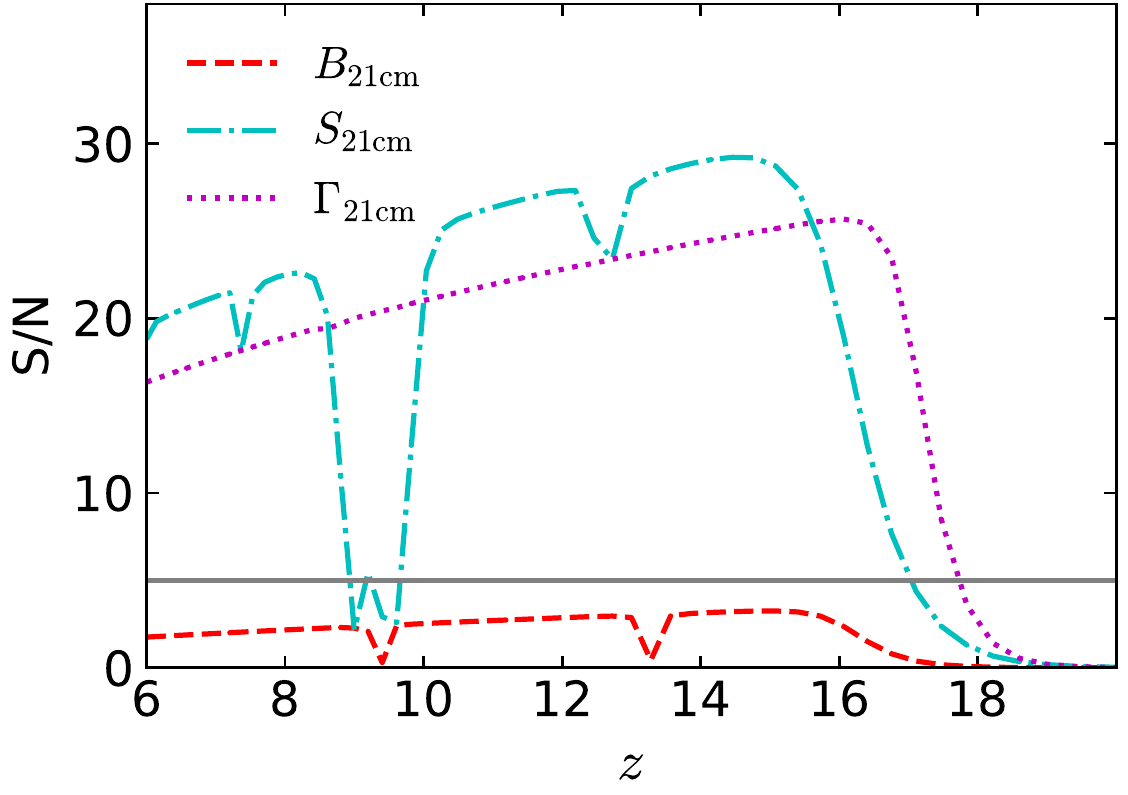}
  \caption{Signal noise  ratios (S/N) of $B_{\rm 21cm}$ (dashed red), $S_{\rm 21cm}$ (dash-dotted cyan) and $\Gamma_{\rm 21cm}$ (dotted magenta) at $k=0.1\,\rm Mpc^{-1}$, predicted for SKA1-low telescope.
  The gray horizontal line denotes the S/N$=5$.
          }
     \label{fig:sn_bis_vz}
\end{figure}
Fig.~\ref{fig:sn_bis_vz} shows the predicted S/N ratios of $B_{\rm 21cm}$, $S_{\rm 21cm}$ and $\Gamma_{\rm 21cm}$ at $k=0.1\,\rm Mpc^{-1}$ for SKA1-low telescope. 
We do not present the results at $k=0.3$ and 1.0 $\rm Mpc^{-1}$, as it is actually hard for the SKA1-low telescope to measure these small scale fluctuations during EoR.
Except some $z$s of phase transition, the S/N ratios of $S_{\rm 21cm}$ is $>20$, and even $\sim 30$ at $z=13.5-15$.
The S/N ratios of $\Gamma_{\rm 21cm}$ is lower, while it is $>20$ at $z=9-17$.
The S/N ratios of $B_{\rm 21cm}$ is $\sim 3$. 
The reason that the S/N ratios of $S_{\rm 21cm}$ and $\Gamma_{\rm 21cm}$ are much higher than $B_{\rm 21cm}$ is because both the $S_{\rm 21cm}$ and $\Gamma_{\rm 21cm}$ are the integration of many $B_{\rm 21cm}$ modes, that increases the detectability to measure the non-Gaussian features of 21-cm signals.

Note that we do not consider the foreground noises, which might significantly reduce the S/N ratios of three-points statistics \citep{Watkinson2021}. 

\section{Conclusions}
\label{sec:concl}
We use semi-numerical simulations (21CMFAST) to study the evolution features of skew spectrum and smoothed skewness of 21-cm signals during the Epoch of Reionization (EoR), and their detectability by SKA1-low telescope.
As a comparison, we also present the results of 21-cm bispectrum.

We run four simulations with the 21CMFAST code, one (L600) is with box length $600\,\rm cMpc$ and grid number $800^3$, and with the normal EoR model. 
Another two are with the same box length and resolution, while one (L600\_fast) with a faster ionization process and the other one (L600\_slow) with a slower ionization process, by increasing and decreasing the star forming efficiency and the halo mass threshold for star formation, respectively. 
The last one (L1200) has the same EoR model with L600, but with a larger box length $1200\,\rm cMpc$.

As the skew spectrum is the cross-correlation of 21-cm images with their square, and smoothed skewness is the 1-D statistics of 21-cm signals, their computing is much easier than the 21-cm bispectrum, while all of them show similar evolution features during EoR, e.g. at $k=0.1$ and $0.3\,\rm Mpc^{-1}$.
They are negative in the Ly$\alpha$ pumping phase, and become positive in the X-ray heating phase.
With the ionization going on, the ionized bubbles result in negative spectra, and at the end of EoR, they are positive again as the non-Gaussian features of 21-cm signals are dominated by the neutral hydrogen islands. 
These evolution features are sensitive to the model of reionization history, while the larger box length simulation L1200 shows consistent results with L600. 

As both the skew spectrum and smoothed skewness are the integration of bispectrum modes \cite[see e.g. ][]{Dai2020, Shimabukuro2016}, their expected S/N ratios for SKA1-low are much higher than that of 21-cm bispectrum, which can be $>20$ at some redshifts.
This denotes that the skew spectrum and smoothed skewness of 21-cm signals should be easier to measure than the 21-cm bispectrum, and can be used to study the physical models of EoR. 

Finally, we summarize that although the 21-cm statistics of skew spectrum and smoothed skewness will miss some modes of 21-cm bispectrum, they are easier to compute and expected with larger S/N ratios measured by the 21-cm experiments e.g. SKA1-low. 
Their measurements in the near future will help to investigate the evolution of EoR. 

\section*{Acknowledgements}
QM is supported by the National SKA Program of China (grant No. 2020SKA0110402), National Natural Science Foundation of China (Grant No. 12263002), Science and Technology Fund of Guizhou Province (Grant No. [2020]1Y020), and GZNU 2019 Special project of training new academics and innovation exploration.
The tools for bibliographic research are offered by the NASA Astrophysics Data Systems and by the JSTOR archive.

\section*{Data Availability}
The simulation data and the post-analysis scripts underlying this article will be shared on reasonable request to the corresponding author.



\bibliographystyle{mnras}
\bibliography{ref} 



\appendix


\bsp	
\label{lastpage}
\end{document}